\documentclass[prl,twocolumn,superscriptaddress]{revtex4-1}
\usepackage{amsmath,amssymb,graphicx}
\usepackage{natbib,hyperref}
\usepackage{bm}

\usepackage{color}

\begin{document}
\author{Areg Ghazaryan}
\affiliation{IST Austria (Institute of Science and Technology Austria), Am Campus 1, 3400 Klosterneuburg, Austria}

\author{Yossi Paltiel}
\affiliation{Applied Physics Department, The Hebrew University of Jerusalem, Bergmann Building, Safra Campus, Jerusalem 91904, Israel}

\author{Mikhail Lemeshko}
\affiliation{IST Austria (Institute of Science and Technology Austria), Am Campus 1, 3400 Klosterneuburg, Austria}

\title{An Analytic  Model of Chiral-Induced Spin Selectivity}


\begin{abstract}
Organic materials are known to feature long spin-diffusion times, originating in a generally small spin-orbit coupling observed in these systems. From that perspective, chiral molecules acting as efficient spin selectors pose a puzzle, that attracted a lot of attention during the recent years. Here we revisit the physical origins of  chiral-induced spin selectivity (CISS), and  propose a simple analytic minimal model to describe it. The model treats a chiral molecule as an anisotropic wire with molecular dipole moments aligned arbitrarily with respect to the wire's axes, and is therefore quite general. Importantly, it shows that helical structure of the molecule is not necessary to observe CISS and other chiral non-helical molecules can also be considered as a potential candidates for CISS effect. We also show that the suggested simple model captures the main characteristics of CISS observed in experiment, without the need for additional constraints employed in the previous studies. The results pave the way for understanding other related physical phenomena where CISS effect plays an essential role.   
\end{abstract}

\maketitle
\section{Introduction}

The main goal and technological challenge of spintronics is to be able to coherently inject, manipulate, and detect spins  in condensed-matter systems~\cite{Zutic2004}. However, despite numerous  spin-based logic devices  proposed during the last decades, the field is still far from being competitive with charge-based architectures~\cite{Awschalom2007,DietlBook}. The limitations partially come from the low level of control of  spin degrees of freedom, which requires both long mean-free path and considerable spin precession due to spin-orbit coupling (SOC). Besides the initial attempts to produce inorganic spintronic devices, organic elements have also been widely explored \cite{Dediu2002,Sanvito2011}. In particular, starting from 1999 \cite{Ray1999} and more actively during the last decade \cite{Gohler2011,Naaman2015} it was shown that chiral molecules can be used as efficient spin signal generators.

Chiral-induced spin selectivity (CISS) denotes the effect in which the electron's spin current acquires a substantial polarization after passing through a monolayer of chiral molecules.  Initially discovered in a double-stranded DNA \cite{Gohler2011},  CISS was later confirmed for other types of molecules \cite{Kettner2015,Mishra2013,Dor2013,Einati2015,Kiran2016,Kettner2018}. As of now, there are several established experimental techniques used to observe CISS. Besides the original photoelectron transmission through a self-assembled monolayer of chiral molecules \cite{Gohler2011}, the CISS effects was also established by spin-specific conduction through chiral molecules, with gold nanoparticles attached to one end of the molecule \cite{Kettner2015,Xie2011}, as well as by the Hall device measurements, where spin-polarization was  accompanied by charge redistribution \cite{Dor2014,Kumar2017}. Currently,  the CISS effect is used as a tool to generate other, quite diverse, physical phenomena \cite{Naaman2015,Michaeli2016,Michaeli2017,Naaman2018,Naaman2019}. As a prominent example, CISS   can be used to generate enantioselectivity, which can have important implications in the biorecognition \cite{Banerjee2018,Naaman2019}. 

While the experimental methods for generating the CISS effect are well established, a comprehensive theoretical approach to this phenomenon is still lacking. Theoretical models usually cluster around two approaches, both of which require additional constraints and assumptions in order to reproduce experimentally observed effects. First type of approaches are based upon  calculating the scattering cross sections within the Born approximation \cite{Yeganeh2009,Medina2012,Varela2013}. In order to account for the observed values of spin polarization, however, it was necessary to increase the magnitude of SOC due to the effective mass renormalization \cite{Yeganeh2009}, sum over incoherent contributions to the  scattering amplitudes from many molecules or many turns of one molecule \cite{Medina2012}, or include inelastic scattering processes~\cite{Varela2013} in the theory. In the second type of approaches, one attaches leads to a chiral molecule in order to calculate its transmission properties \cite{Guo2012,Gutierrez2012,Gutierrez2013,Guo2014,Matityahu2016,Michaeli2015,Yang2019,Dalum2019, Geyer2019}. Similarly, addition of extra terms corresponding to dissipation \cite{Guo2012}, next-nearest-neighbor hopping \cite{Guo2014}, or a molecular axis aligned dipolar field \cite{Michaeli2015} was required in order to obtain considerable spin polarization in such transport calculations. A different modeling of CISS was given recently, by exploring the idea that initially not all possible states of the electron with the same energy are excited and the effect of SOC is enhanced due to the degeneracies of the excited states of the molecule \cite{Dalum2019}. In contrast to other works, the effect of the SOC in the substrate for generation of CISS effect was also explored \cite{Gersten2013}. Finally the role of electron-electron correlations in the molecule for CISS effect has also been addressed lately \cite{Fransson2019}.  

By now it is theoretically usually agreed that the necessary ingredients for observing CISS are (i) molecular chirality and (ii) a considerable amount of SOC. At the same time, the actual magnitude of SOC \cite{Varela2016}, as well as  the importance and the physical meaning of the extra terms listed above is highly debated. Therefore, it is of high importance to develop a simple and general model that  captures the main physics of CISS with a relatively small number of  parameters. Such a theory would allow one to understand the role played in CISS by each ingredient and to make direct predictions for future experiments. Besides, it should be pointed out that all previous theoretical models for chiral molecules on a substrate \cite{Eremko2013,Gersten2013,Guo2012,Guo2014,Gutierrez2012,Gutierrez2013,Matityahu2016,Medina2012,Michaeli2015,Yeganeh2009,Varela2013,Geyer2019,Dalum2019,Varela2016} employ as a model for the molecule in a form of helix (see Fig.~\ref{fig:PictorialFigure}). The notion, that for the observation of the CISS effect the necessary ingredient is chirality and not helicity, have not been stressed considerably in the literature, except for the initial papers for the gas phase \cite{Blum1989,Blum1990,Blum1998}. While the effective electrostatic potential experienced by the electrons turns out to be helical in general, the modeling for the molecule using a potential as a helix or spiral is not a necessary ingredient to obtain CISS effect.      

Here we propose an analytically tractable minimal model, which captures the main characteristics of CISS. Motivated by the microwave spectroscopy measurements of chiral molecules~\cite{Patterson2013}, we model the molecule as a anisotropic wire with the dipole field which is not aligned along any specific molecular axis, see Fig.~\ref{fig:PictorialFigure}. Our theory is able to  reproduce the  CISS effect observed in experiment using realistic material parameters and without introducing any extra terms into the model Hamiltonian. We believe that our approach can be used as a starting point to explore a variety of experimental measurements relying on the CISS, such as observation of unconventional triplet pairing superconductivity induced by chiral molecules in  $s-$wave superconductors  \cite{Alpern2016}, enantioselectivity using an achiral magnetic substrate \cite{Banerjee2018}, observed correlations between  charge and spin separation in chiral molecules \cite{Kumar2017}, and other related phenomena. Besides our theory suggests, that other chiral but non-helical molecules can also be used to observe the CISS in the experiment.

\section{Theoretical model}
\begin{figure}
\includegraphics[width=8cm]{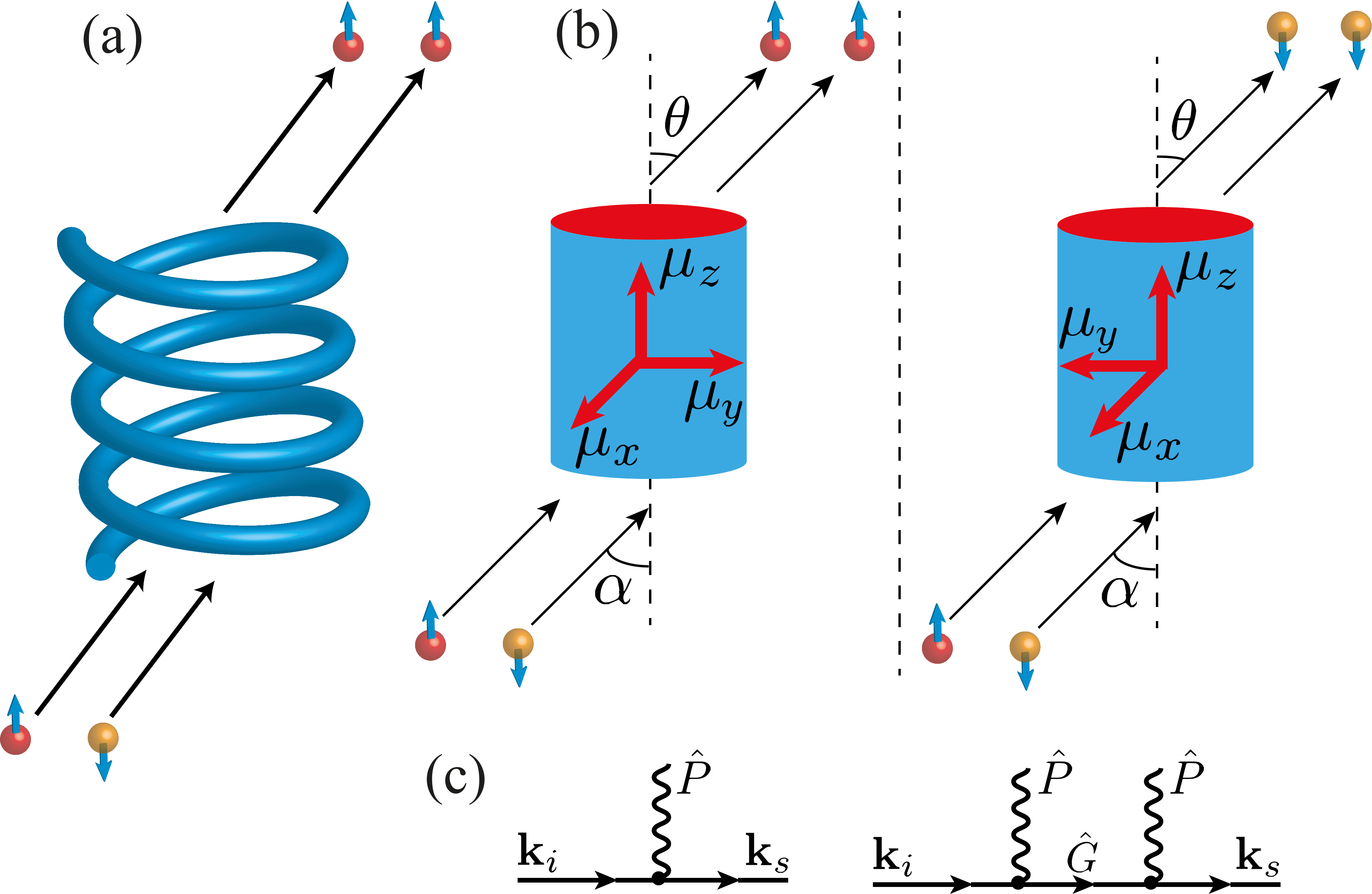}
\caption{\label{fig:PictorialFigure} Pictorial representation of the molecule: (a)~As a helix used in the previous studies and (b)~As a finite quantum wire employed in this study. Implementation of two enantiomers with finite quantum wire potential is also shown. Electrons are depicted as red and yellow spheres, where color and arrows is related to the spin of the particles. $\mu_x,\mu_y$ and $\mu_z$ are different components of the dipole field of the molecule. (c) Graphical representation of the scattering events considered in (\ref{BornSeries}).
}\end{figure}

We start from a general  Hamiltonian describing the system of Fig.~\ref{fig:PictorialFigure} (b):
\begin{equation}
\mathcal{H}=-\frac{\hbar^2\nabla^2}{2m}+V_D(\mathbf{r})+V_\mathrm{SOC}(\mathbf{r}),
\end{equation}
Here $V_D(\mathbf{r})$ is the interaction of the electron with the electric field of molecular dipoles, which are confined due to the wire geometry and $V_\mathrm{SOC}(\mathbf{r})$ describes the SOC due to that potential.  We model the molecule as a finite anisotropic quantum wire. The molecular dipole moment with components $\boldsymbol{\mu}=(\mu_x,\mu_y,\mu_z)$ can point in an arbitrary direction with respect to the molecular axes. Therefore, the dipolar part of the Hamiltonian will be
\begin{equation}
\label{eq:Dip}
V_D(\mathbf{r})=\left(eE_x x+ eE_y y + eE_z z\right)e^{-\frac{\xi^2_1x^2+\xi^2_2y^2+\xi^2_3z^2}{2}},
\end{equation}
where $e$ is the charge of the electron and the electric field components $E_x,E_y$ and $E_z$ are produced by the corresponding dipole components. $\xi_1$, $\xi_2$ and $\xi_3$ determine the sizes of the aniostropic wire. We evaluate $E_i$  as the value of an electric field at the center of the dipole  $\mu_i$, modeled as a two point charges. That is,  $E_i=8\mu_i/l_i^3$, where $l_i$ is the length of the molecule in the corresponding direction. We take the length to be defined as the full width half maximum of the Gaussian profile of the dipole field ($l_i\xi_i=2\sqrt{2\ln2}$). The SOC is calculated as the Rashba SOC due to the dipolar field as
\begin{equation}
V_\mathrm{SOC}(\mathbf{r})=-i\alpha_\mathrm{SOC}\boldsymbol{\sigma}\cdot\left[\nabla V_D(\mathbf{r})\times\nabla\right]
\label{SOCPot}
\end{equation}
As   discussed below,  the actual value of SOC in chiral molecules is quite different from the SOC parameter for a free electron, $\alpha_\mathrm{SOC}=\hbar^2/4m^2c^2$.

We consider the CISS effect from the perspective of an incident electron scattering from the potential $P(\mathbf{r})=V_D(\mathbf{r})+V_\mathrm{SOC}(\mathbf{r})$ of Eqs.~\eqref{eq:Dip} and~\eqref{SOCPot}. We start from a free particle in the initial state $\langle \mathbf{r},s|\psi_0\rangle=\psi_0(\mathbf{r})\chi_{m_s}=e^{i\mathbf{k}_i\mathbf{r}}\chi_{m_s}$, where $\mathbf{k}_i=k(\sin\alpha\cos\beta,\sin\alpha\sin\beta,\cos\alpha)$,  $\alpha$ and $\beta$ determine the polar and azimuthal angle of the incident electron, and $\chi_{m_s}$ is the incident spin state. Within the second-order Born approximation, the scattering state is given by (see Fig.~\ref{fig:PictorialFigure} (c)): 
\begin{equation}
|\psi\rangle=|\psi_0\rangle+\hat{G}\hat{P}|\psi_0\rangle+\hat{G}\hat{P}\hat{G}\hat{P}|\psi_0\rangle,
\label{BornSeries}
\end{equation}
where $\hat{G}$ is the Green's function for the free particle $\langle \mathbf{r}|\hat{G}|\mathbf{r}^\prime\rangle=G(\mathbf{r},\mathbf{r}^\prime)=-e^{ik\left|\mathbf{r}-\mathbf{r}^\prime\right|}/4\pi\left|\mathbf{r}-\mathbf{r}^\prime\right|$. We are interested in the state of the electron far away from molecule, so we can approximate the leftmost Green's function in Eq.~(\ref{BornSeries}) by its well known asymptotic form, $G(\mathbf{r},\mathbf{r}^\prime)=-\left(e^{i k r}/4\pi r\right)e^{-i\mathbf{k}_s\mathbf{r}}$, where $\mathbf{k}_s=k(\sin\theta\cos\tau,\sin\theta\sin\tau,\cos\theta)$, and $\theta$ and $\tau$ determine the polar and azimuthal angles of the scattered electron. We consider only elastic (energy-conserving) scattering of the electron on the molecule. Since spin polarization is zero in the up to first order (see below), we need to calculate the second order correction as well. Unfortunately, this cannot be done analytically for the case of free particle Green's function. Therefore, we consider additionally the potential $V_w(\mathbf{r})=\left(\hbar^2/2m\right)\left(x^2/a^4_x+y^2/a^4_y\right)$, which spatially confines the scattering event in $xy$ plane ($a_x$ and $a_y$ define the characteristic lengths of the potential). While, in the current treatment it is just a mathematical tool to make the second order scattering analytically tractable, it can be justified also physically as a wave packet width of the incoming electron or finite size of the sample. Using the complete basis of the eigenstates of this $x-y$ potential, $\phi_{n_1,n_2}(x,y)$, where $n_1$ and $n_2$ are quantum numbers of two separate harmonic oscillators, we can insert the completeness relation into the integrals of (\ref{BornSeries}), which greatly simplifies the calculation. The summations are truncated at finite values of $n_1$ and $n_2$. The cutoff values of $n_1$,$n_2$ and characteristic lengths $a_x$, $a_y$ are determined from the condition, that up to first order the outgoing current should be comparable to the current obtained with the free particle Green's function treatment.
After evaluating all the integrals analytically, we arrive at the final expression of the asymptotic form of the wave function
\begin{equation}
\psi(\mathbf{r},s)=\left[\psi_0(\mathbf{r})+\frac{e^{i k r}}{r}\hat{f}(\mathbf{k}_i,\mathbf{k}_s)\right]\chi_{m_s},
\end{equation}
where $\hat{f}(\mathbf{k}_i,\mathbf{k}_s)$ is the scattering amplitude, which is an operator in the spin space \cite{Taylor1972,Sitenko1991}. Quite generally $\hat{f}(\mathbf{k}_i,\mathbf{k}_s)=\sum_{m=0}^{3}f_m(\mathbf{k}_i,\mathbf{k}_s)\sigma_i$, where $\sigma_i$ for $i=1,2,3$ are Pauli matrices, $\sigma_0$ is the identity matrix and $f_i(\mathbf{k}_i,\mathbf{k}_s)$ are complex functions. Once the scattering amplitude is computed, the polarization of the outgoing beam in the direction $a=\{x,y,z\}$ can be evaluated as $P_a=\mathrm{Tr}(\boldsymbol{\sigma}_a\boldsymbol{\rho}^\prime(\mathbf{k}_i,\mathbf{k}_s))/\mathrm{Tr}(\boldsymbol{\rho}^\prime(\mathbf{k}_i,\mathbf{k}_s))$, where $\boldsymbol{\rho}^\prime(\mathbf{k}_i,\mathbf{k}_s)=\hat{f}(\mathbf{k}_i,\mathbf{k}_s)\boldsymbol{\rho}_0\hat{f}^\dagger(\mathbf{k}_i,\mathbf{k}_s)$ is the denisty matrix of the outgoing beam and $\boldsymbol{\rho}_0$ is the density matrix of incoming beam. In particular, for a  non-polarized beam $\boldsymbol{\rho}_0=(1/2)\mathbb{I}$. It can be shown that \cite{Medina2012}
\begin{align}
P_a=&\frac{1}{\sum_{m=0}^{3}\left|f_m(\mathbf{k}_i,\mathbf{k}_s)\right|^2}\left(2\mathrm{Re}\left[f_0(\mathbf{k}_i,\mathbf{k}_s)f^\ast_a(\mathbf{k}_i,\mathbf{k}_s)\right]-\right. \nonumber \\
&\hspace{80pt}\left.2\mathrm{Re}\left[if_l(\mathbf{k}_i,\mathbf{k}_s)f^\ast_p(\mathbf{k}_i,\mathbf{k}_s)\right]\right),
\label{PolFormula}
\end{align}
where indices $(a,l,p)$ form a cyclic order. It is fairly easy to see from this relation that spin polarization up to first order should be zero. Taking into account that up to first order $\hat{f}(\mathbf{k}_i,\mathbf{k}_s)=-\frac{m}{2\pi\hbar^2}\int d\mathbf{r} e^{-i\mathbf{k}_s\mathbf{r}}\left(V_D(\mathbf{r})+V_\mathrm{SOC}(\mathbf{r})\right)e^{i\mathbf{k}_i\mathbf{r}}$ and that $V_D(-\mathbf{r})=-V_D(\mathbf{r})$, $V_\mathrm{SOC}(-\mathbf{r})=V_\mathrm{SOC}(\mathbf{r})$, it is easy to show that $f^\ast_0(\mathbf{k}_i,\mathbf{k}_s)=-f_0(\mathbf{k}_i,\mathbf{k}_s)$ and $f^\ast_m(\mathbf{k}_i,\mathbf{k}_s)=f_m(\mathbf{k}_i,\mathbf{k}_s)$ $m=1,2,3$. This directly shows that for the chosen potential the polarization (\ref{PolFormula}) in any direction is always zero up to first order of perturbation theory. Therefore, finite spin polarization can only be obtained considering the scattering up to second order of perturbation theory.

\section{Results and discussion}
In what follows, we present the results of the spin polarization in $z$ direction for different values of the model parameters. In the current calculation we choose $l_x=0.8\,\mathrm{nm}$, $l_y=2.0\,\mathrm{nm}$ and $l_z=10.0\,\mathrm{nm}$ \cite{Nguyen2019}. For the magnitudes of the molecular dipole moments we use $\mu_x=2.4\,\mathrm{D}$, $\mu_y=5.0\,\mathrm{D}$, $\mu_z=1.8\,\mathrm{D}$ as representative values. These values are comparable to the experimentally measured values for 1,2-propanediol or for molecules with similar structure \cite{Lovas2009,Patterson2013}. As noted above, we calculate scattering terms up to second order perturbation theory. For this regime we can both observe finite polarization and make the model analytically tractable. 

By analogy to the previous theoretical studies~\cite{Varela2016}, the amplitude of SOC requires special consideration. It is now well established that free electron SOC, $\alpha=\hbar^2/4m^2c^2$, is too small to account for CISS effect. In fact the situation for chiral molecules is quite similar to the case of graphene and carbon nanotubes, where the issue of SOC  has been actively studied before \cite{Ando2000,Min2006,Huertas2006,Izumida2009}. Generally, there are three types of SOC which can be realized by combining the SOC of carbon atoms and the overlap of the orbitals at the adjacent sites of the lattice (motion of the electrons in the lattice). The first type is the intrisic SOC due to the carbon atoms and overlap of the $\sigma$ orbitals of nearest-neighbors. The second type, the Rashba SOC, is a combination of the Stark effect, SOC of atoms, and $\sigma$ orbital overlap. Generally, the effective SOC of $\pi$ bands due to these two mechanisms is considered to be small, which limits  potential applications of graphene in spintronics. Finally, the third mechanism is due to the curvature effect, which induces hopping between $\pi$ and $\sigma$ bands. In conjunction with SOC of carbon atoms this generates an effective SOC for $\pi$ bands for carbon nanotubes which is at least an order of magnitude larger than the SOC calculated for graphene. This observation was confirmed by several theoretical \cite{Ando2000,Huertas2006,Izumida2009} and experimental \cite{Kuemmeth2008,Jhang2010,Steele2013} studies. In fact the large amplitude of SOC observed in \cite{Steele2013} is hard to account for even considering curvature effects. Similar reasoning also applies to chiral molecules \cite{Varela2016}, which justifies using a renormalized value of the SOC magnitude in the calculations. In order to determine the actual value of SOC in the current calculation we consider the same model of the molecule, but take it to be infinite in $z$ direction. By numerically calculating the energy spectrum of that system we find that when the bare electron SOC is renormalized by a factor $10^7$ the resulting energy splitting is approximately $40-80\,\mathrm{meV}$. There were no direct experimental measurements of the energy splitting for the chiral molecule due to the SOC. As was noted above, these type of measurements have been performed for carbon nanotubes \cite{Kuemmeth2008,Jhang2010,Steele2013} and a splitting in the range of a few meV was observed. Recently, singlet-triplet splitting for injected electron from ferromagnet into chiral monolayer was measured experimentally through Kelvin-probe force microscopy and the value of 30 meV was found \cite{Ghosh2020}. While this value is a consequence of not only SOC, but also electron exchange interaction between substrate and the molecule, we correlate the SOC splitting in the current model with this value. This is partially justified, since the effect of the substrate in the current model is taken into account only phenomenologically. Therefore, we apply a similar $10^7$ factor renormalization of the SOC magnitude in Eq.~(\ref{SOCPot}) in order to obtain results comparable with  experimental observations.

\begin{figure}
\includegraphics[width=9cm]{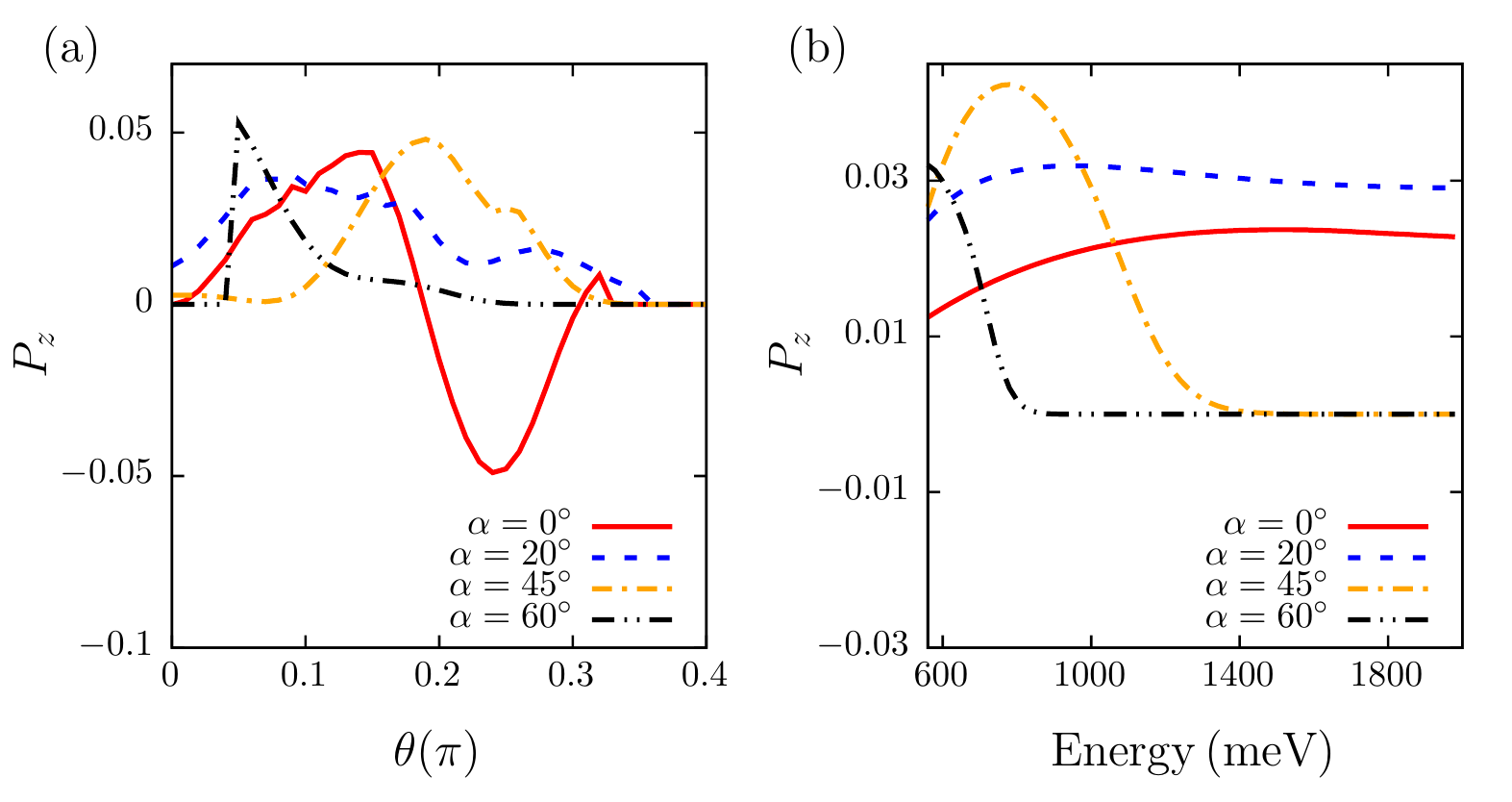}
\caption{\label{fig:PolDepAngEn} Dependence of the outgoing beam spin polarization in the $z$ direction (a) on the polar angle $\theta$ and (b) on the incoming electron energy when integrated over $\theta$ in the range $\left[0-\pi/2\right]$, for different angles of incoming polar angle $\alpha$. The incoming electron energy is $1000\,\mathrm{meV}$ in (a); the results are integrated over both incoming and outgoing azimuthal angles $\beta$ and $\tau$. 
}\end{figure}

Another aspect of the measurement worth further consideration is integration of the input and output channels over the angles. While integration over azimuthal angles for both incoming and outgoing cases is justified,  integration over polar angle is more subtle. In the current study, we both show the results for the case of specific polar angles $\alpha$ and $\theta$ as well as results when all the angles are integrated out. We chose the range of integration for azimuthal angles from $0$ to $2\pi$ and for polar angles from $0$ to $\pi/2$. It should be noted that the incoming polar angle $\alpha$ is related to the orientation of the molecule with respect to the surface of the substrate in the experiment. Therefore, ideally integration range of $\alpha$ should be controlled by the orientation of the molecule and the range $0$ to $\pi/2$ is not fully justified. Since this issue is also related to material specifics of the substrate (probability distribution of the outgoing electrons) and also imperfections of the surface growth, we do not take them into account in the current model study. It should be noted that for perfectly aligned molecule the integration of the polar angle in the range of $0$ to $\pi/2$ captures all the electrons entering into and living from molecular monolayer. Therefore, for this ideal arrangement this range of integration captures all the transferred electrons which mimics the situation usually observed in the experiment.

Fig.~\ref{fig:PolDepAngEn} (a) shows the $\theta-$dependence of  spin polarization  in the $z$ direction  in the outgoing beam, for the incoming electron energy of $1000\,\mathrm{meV}$. As one can see from the figure, the  obtained polarization reaches up to several percent. While these numbers are smaller than the ones observed in the experiment, by modifying the parameter values it is possible to get results comparable to the experiment. Since our goal here is to demonstrate that chiral non-helical system can act as a spin polarizer and in the mean time use analytically tractable model, we refrain from such parameter tuning. As can be seen from the figure polarization shows oscillations with the outgoing angle. While this would suggest that overall integrated polarization should be small, this viewpoint is misleading. The reason for that is physically the experimental setup does no integrate polarization, but rather the current. Therefore, while in some directions polarization can be quite large, the contribution of the current in that direction to the overall polarization can be minor. When calculating polarization in the directions where the current is several tens of magnitude smaller than the original one, we just put the polarization result to be zero (this explains the strict zero result observed in the figure).

Fig.~\ref{fig:PolDepAngEn} (b) shows the dependence of  spin polarization in $z$ direction of the outgoing beam on the incoming electron energy, when integrated over outgoing angle $\theta$ in the range $\left[0, \pi/2\right]$. As can be seen from the figure the polarization is relatively constant with respect to energy for small incoming angles $\alpha$. While for larger values of $\alpha$ the polarization shows pronounced peaks for some energies it quickly drops to zero for larger values of energy. Physical consequence of this observation is well known in the experiment. In order to get stable polarization results the molecules should be aligned perpendicular to the substrate, which means reducing incoming angle $\alpha$. This is justified also for stronger attachment of the molecule to the substrate and for aligned organization of the molecules in the monolayer, which is known to affect the efficiency of CISS effect \cite{Aqua2003,Ray2006}. We have checked that the obtained results are reversed when flipping to the other enantiomer, which in the current model can be done by adjusting the relative direction of the dipolar field with respect to the anisotropic wire potential (see Fig.~\ref{fig:PictorialFigure} (b)). Also we have confirmed that the effect disappears, if we make the wire potential circular or align the dipolar field  along one of the wire's axes. Therefore, the observed spin polarization is due to the chirality of the system and both the non-circular wire potential and arbitrary aligned dipolar field are essential ingredients for observing CISS.

\begin{figure}
\begin{tabular}{cc}
\includegraphics[width=4.2cm]{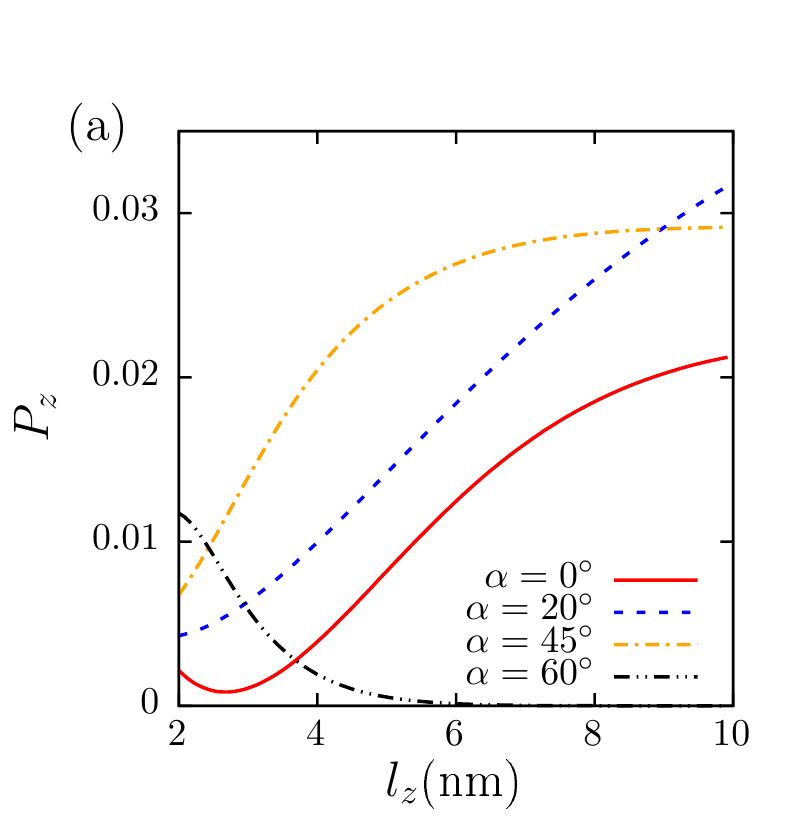} &
\includegraphics[width=4.2cm]{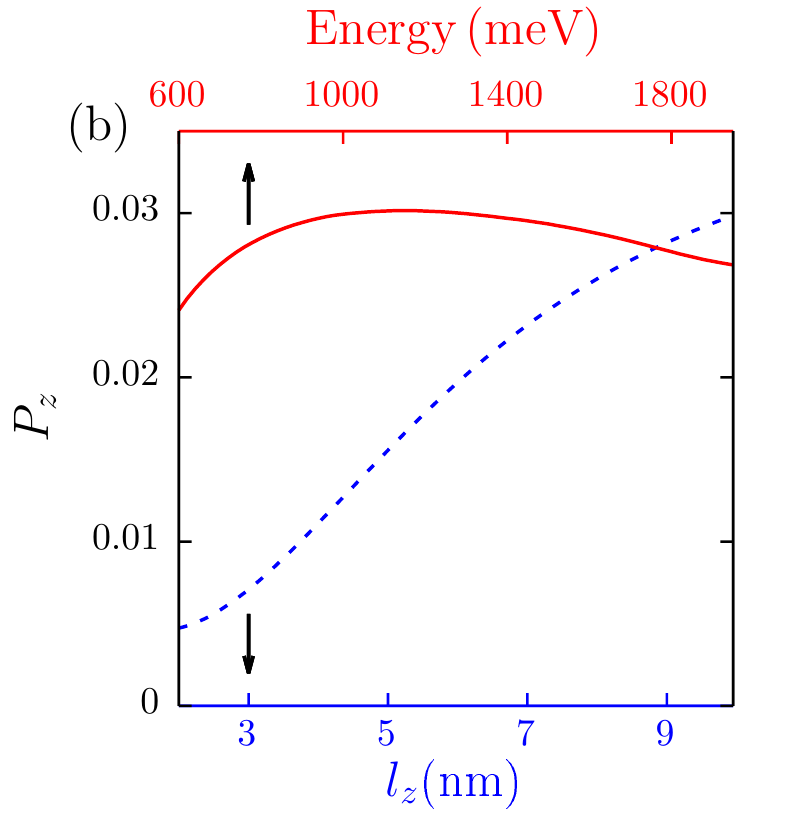}
\end{tabular}
\caption{\label{fig:PolDepLzInt} (a) Dependence of the outgoing beam spin polarization in the $z$ direction on the length of the molecule $l_z$, for different angles of incoming polar angle, $\alpha$. The results are integrated over outgoing polar angle $\theta$ and in the incoming and outgoing azimuthal angles $\beta$ and $\tau$. (b) Dependence of the fully integrated outgoing beam spin polarization in the $z$ direction on the length of the molecule $l_z$ and energy of incoming electrons. The integration range for polar angles is again $\left[0, \pi/2\right]$ and for azimuthal angles $\left[0, 2\pi\right]$. The electron energy is $1000\,\mathrm{meV}$ for the length dependence curves.}\end{figure}

It was known from early on \cite{Gohler2011} that increasing the length of the molecule decreases the amplitude of outgoing current due to backscattering and electron capture~\cite{Markus2013}, however it increases the observed spin polarization. This has been already confirmed experimentally for different systems \cite{Kumar2013,Kettner2015,Ghosh2019}. To test that feature, in Fig.~\ref{fig:PolDepLzInt} (a) the dependence of the polarization on the length of the molecule is shown for different values of incoming polar angle $\alpha$. As in Fig.~\ref{fig:PolDepAngEn} (b) the results are integrated over azimuthal angles and  the outgoing polar angle $\theta$ in the range of $\left[0, \pi/2\right]$. For small incoming polar angles the polarization indeed increases with the length of the molecule, taking into account also the fact that electric field in that direction decreases. The trend changes only for larger incoming angles (for $\alpha=60^\circ$ in the figure), although in this case the outgoing current is fairly small.


Finally Fig.~\ref{fig:PolDepLzInt} (b) shows the dependence of the spin polarization in $z$ direction on the energy of electron and the length of the molecule, when all angles are integrated out. As in the previous cases the azimuthal (polar) angles are integrated in the range of $0$ to $2\pi$ ($0$ to $\pi/2$). As can be seen from the figure, the results are quite similar to the case with $\alpha=0$, since predominant outgoing current is produced by these electrons. Again the polarization is almost constant with the change of energy of electron and increases with the length of the molecule, which is qualitatively similar to the trends observed in the experiment. 

\section{Conclusions}
In conclusion, in this work we have constructed a minimal model of a chiral molecule, which captures the main characteristics of the CISS effect. In particular, in comparison to previous studies, our model does not assume helical molecular structure,  but models the molecule as an anisotropic potential in combination with a electric dipole field. In such a setting, all the terms in the scattering theory up to second order can be evaluated analytically. The role of dipole field is crucial in our model, since the chirality of the system is determined by the mutual orientation of the electric dipole moment and anisotropic wire potential. The current model ignores substrate effect considering it only as a source of unpolarized electron current. We have shown that the current model produces considerable spin polarization. It is also demonstrated that not only the chirality of the molecule, but also the alignment of the molecule with respect to substrate plays an important role in the overall spin polarization observed in the outgoing current. Being quite general and analytically solvable, the model can be used to describe other related physical phenomena where chirality of the molecule plays an essential role. Additionally, we hope that this will stimulate further experimental studies of CISS effect for chiral molecules which does not posses helical structure. 

Finally it should be noted that, the proposed model being minimal cannot account for the effects of specific atomic potentials of the molecules on the nature of CISS. This type of refinements can be included by considering an extended and more complex model of the molecule, which can shed light on the specific characteristics of the constituent atoms responsible for CISS. These issues will be addressed in our future studies.  

\section{Acknowledgments}
A.G. is grateful to Artem Volosniev for valuable discussions. A.G.~acknowledges support from the European Union’s Horizon 2020 research and innovation program under the Marie Sk\l{}odowska-Curie grant agreement No 754411. Y.P.~acknowledges support from the Volkswagen Foundation (No. VW 88 367), and the Israel Ministry of Science (MOS). M.L.~acknowledges support by the Austrian Science Fund (FWF), under project No.~P29902-N27, and by the European Research Council (ERC) Starting Grant No.~801770 (ANGULON).

\bibliography{chiralmolbib}
\bibliographystyle{apsrev4-1}

\end{document}